\documentstyle[preprint,prb,aps,epsf]{revtex}
\def\inseps#1#2{\def\epsfsize##1##2{#2##1} \centerline{\epsfbox{#1}}}
\def\rect#1#2{{\vcenter{\vbox{\hrule height.3pt
            \hbox{\vrule width.3pt height#2truecm \kern#1truecm
            \vrule width.3pt}
            \hrule height.3pt}}}}
\def\square{\rect{0.15}{0.15}}
\def\blacksquare{\vrule height 0.15truecm width 0.15 truecm depth -0.1ex}
\begin{document}
\draft
\input{psfig}
\title{Persistence exponent in superantiferromagnetic quenching}
\author{E.N.M. Cirillo$^1$, G. Gonnella$^2$\footnote{Corresponding author}
and S. Stramaglia$^3$}
\address{$^1$Universit\'e Paris Sud, Mathematique, \\
B\^atiment 425, 91405 Orsay Cedex, France. E\_mail:
Emilio.Cirillo@math.u-psud.fr} 
\address{$^2$Istituto Nazionale per la Fisica della Materia {\rm and}
Dipartimento di Fisica dell'Universit\`a di Bari, 
via Amendola
173, 70126 Bari, Italy. E\_mail: Giuseppe.Gonnella@ba.infn.it}
\address{$^3$Istituto Elaborazione Segnali e Immagini - Consiglio Nazionale
delle Ricerche, via Amendola 166/5, 70126 Bari, Italy.
E\_mail: sebino@iesi.ba.cnr.it}
\date{\today}
\maketitle
\begin{abstract}
We measure the persistence exponent in a phase separating two-dimensional
spin system with non-conserved dynamics
quenched in a region with four coexisting stripe phases.
The system is an Ising model with nearest neighbor, next-to-the-nearest
neighbor and plaquette interactions. Due the particular nature
of the ground states, the order parameter is defined in terms of 
blocks of spins. Our estimate of the persistence exponent,
$\theta=0.42$, differs from those of the two-dimensional 
Ising and four state Potts models. 
Our procedure allows the study of persistence properties
also at finite temperature $T$:
our results are compatible with the hypothesis that $\theta$
does not depend on $T$ below the critical point.
\end{abstract}


\newpage

\section{Introduction}

When a system is quenched from a disordered phase into a multiphase
coexistence region, ordered  domains form randomly and grow in a 
self-similar way [\onlinecite{BBB}]. The kinetics of coarsening domains is
described by the algebraic growth $L(t) \sim t^z$ where $L(t)$
is the characteristic length of  domains at time $t$ and $z$ is an
universal exponent not depending on the dimensionality of the system nor 
on the final temperature of the quenching. 
In most models with
non-conserved order parameter $z=1/2$, 
while in scalar models with conserved
order parameters $z=1/3$ [\onlinecite{B94}]. 
There is a general agreement on this scenario
and the values of the exponent $z$ are used to classify the growth 
regimes in experimental systems. 

More recently, new dynamical exponents have been considered also with
the hope to better characterize the process of phase separation. 
However, the universal character of these new exponents is under debate
and the determination of their values in particular cases can be useful to
clarify their nature. 

A new exponent introduced by Derrida et al. [\onlinecite{DBG}]
is related to the
persistence probability $p(t)$ defined as the probability that
the local order parameter at a given point $x$ has never changed  
since the initial time. In spinodal decomposition problems this probability
was first considered for the Ising model with Glauber dynamics at $T=0$.
Here $p(t)$ is the fraction of spins that have never flipped from the beginning
of the process. This probability decays as $p(t) \sim t^{-\theta}$ and 
$\theta$ is called the persistence exponent. In $d=1$ $\theta$ 
has been calculated
exactly for the Potts model with non-conserved dynamics
[\onlinecite{DHP}], for the Ising model
$\theta=3/8$. In higher dimension the evaluation of  $\theta$
is based on Monte Carlo simulations [\onlinecite{S}]. 
For example in the two-dimensional Ising model $\theta=0.22$
and different values have been calculated for other dimensions and Potts 
model [\onlinecite{DOS},\onlinecite{H}]. 
All these results concern the case $T=0$.

At temperatures different from zero, fluctuations induce an exponential
decay of the probability $p(t)$ and the definition of the persistence 
exponent is more subtle. A method has been proposed  
[\onlinecite{D}]
to evaluate $\theta$ at $T\ne0$ based on the comparison of two
different systems, one starting from a random and the other from 
a ground state,
evolving with the same noise. The spin flips in the ordered state
represent the effects of thermal fluctuations and are to be subtracted
to the spin flips of the phase separating 
 system in order to measure  $\theta$.
This method has been applied to Potts model with non-conserved dynamics
[\onlinecite{HH}].
\par
In this paper we calculate the persistence exponent in a two-dimensional
spin system quenched in a region with four coexisting stripe phases.
The system is an Ising model with nearest neighbor, next-to-the-nearest
neighbor and plaquette interactions. Due the particular nature
of the ground states, the order parameter is defined in terms of 
blocks of spins. We study the persistence properties of these blocks 
and find a value of the persistence exponent different from that of
Ising and Potts model. 
The method has been  used also for calculating the  persistence exponent
at $T\ne 0$. Our results are compatible with the hypothesis that $\theta$
does not change with the temperature.
\par
Section \ref{sec:modello}
in the paper is devoted to the description of the model
and of the block variables used for describing the different phases.
In Section \ref{sec:risultati}
our results will be presented for various temperatures.
Some conclusions will follow.

\section{Model and methods}
\label{sec:modello}
The model we consider is the Ising version
of the two-dimensional isotropic eight vertex model [\onlinecite{Bax}], 
with hamiltonian $H$ given by
\begin{equation}
-\beta H = J_1 \sum_{<ij>} s_i s_j + J_2\sum_{<<ij>>} s_i s_j
          +J_3 \sum _{[i,j,k,l]} s_i s_j s_k s_l,
\label{eq:ham}
\end{equation}
where $s_i$ are Ising spins on a two-dimensional square lattice and the sums 
are respectively on nearest neighbor pairs of spins, next to the nearest 
neighbor pairs and plaquettes. Periodic
boundary conditions are always assumed.
At $J_2 < 0, |J_1| < 2 |J_2|$ and $J_3$ small
a critical line separates the paramagnetic phase
from a region where four phases corresponding to ground states
with alternate plus and minus columns or rows coexist [\onlinecite{Bax}].
The system is prepared 
in  a completely disordered
 configuration and the persistent behavior is studied 
in  the stripe-ordered phase coexistence region 
 by numerical simulations both at 
zero and finite temperature
with heat-bath single-spin-flip dynamics [\onlinecite{Gla}]. 
\par
The study of the  evolution of the local order parameter 
in the phase separation process has to take
into account the particular structure of the ground states. 
In  Ising  spin models the persistent fraction  $p(t)$ is generally
defined as the number of spins that never changed sign in the interval $[0,t]$ 
normalized with respect to the volume of the system. 
Here we have to consider the persistence of the local order parameter
and so we must label each site as belonging to one of 
the four possible lamellar ground states. 
\par
The definition of the local order parameter
is based on a  block method that  has already been used  in 
[\onlinecite{CGS1},\onlinecite{CGS2}] for studying the growth
of the stripe  domains of model (\ref{eq:ham}). Let us focus on the site
$x$ of the lattice at time $t$: we consider a $(2l+1)\times (2l+1)$
square $S_l$ with $l$ a positive integer number and we compare
the configuration of the system at time $t$ inside $S_l$ with
the possible four ground states. If one of the four ground states
agrees with the configuration in $S_l$ except a number of
spins less than or equal to a threshold value
$K$ we say that the site $x$ at time $t$ belongs to this 
ground state and we assign it the color $q=1,...,4$; 
otherwise we say that the site is 
a defect and we do not assign it any color.
All the simulations that will be discussed throughout the paper
have been performed by choosing $K$ such that 
$K/|S_l|=K/(2l+1)^2\sim 0.1$,
obviously we have checked that our results do not depend on the choice
of $K$. In [\onlinecite{CGS1},\onlinecite{CGS2}] we have measured by this
procedure the growth exponent and shown that the average size of domains
growths as $R(t)\sim t^z$ where $z=1/2$ or $z=1/3$ in the non-conserved
and conserved cases respectively. 
\par
We use the above definitions to measure 
the fraction $Q(t)$ of the number of sites  
not changing their color:
in our procedure 
we let the system evolve in the interval $[0,t_0]$,
so that domains form,
we label all the spins at time $t_0$ and then we count the number
of spins that do not change color during the further evolution.
This method  resembles the ``block persistence" approach introduced in 
[\onlinecite{CS}] to measure persistence exponents in the Ising
model at finite temperature. We 
remark that in our problem  the introduction of blocks is suggested by the character of the local order parameter  
not defined on a single site.
\par
The definition of the  persistence exponent at
 finite temperatures requires some observations.
At zero temperature
a spin flip occurs  only when the site of the spin is reached
by the boundary of a growing domain and 
the number of persistent spins decreases
as  $p(t)\sim t^{-\theta}$. At finite temperature 
a spin can flip randomly due to
thermal disorder. That is, in the probability of one spin flip there
is a constant term, not depending on time; this suggests the  exponential 
correction $p(t)\sim t^{-\theta}\times\mathrm{e}^{-\lambda t}$ [\onlinecite{sirenetta}].
\par
In the block method the effect of  spin
flip due to thermal disorder can be eliminated in a natural way.
When the temperature increases 
larger values of the side length $2l+1$ of the block used to 
label the spins can be used.
Indeed, to destroy the local order inside a block
$S_l$ it is necessary that the number of thermal-due 
flips exceed $K$. This means that the exponential
correction due to thermal effect can be reduced by considering
a sufficiently large block $S_l$, so that a thermal destruction
of the order inside $S_l$ is unlikely.
\par
Let us describe the following example:
we initialise model (\ref{eq:ham}) 
in an ordered phase and let it evolve 
at finite temperature and parameters $J_1=J_3=0$ and $J_2=-0.5$. After $1000$ 
full sweeps of the grid the system is assumed to be in thermal equilibrium.
Starting from the so obtained configuration, shown in 
Fig. \ref{fig:config},
we apply our block procedure to label all the sites of the
lattice. Since all defects are thermal due,
all the lattice sites should be recognized as belonging to the same phase.
In the four pictures of Fig. \ref{fig:filtri}
black squares represent sites not recognized as
belonging to the ordered phase for blocks $S_l$ $2\times 2$ 
[\onlinecite{osservazione}], 
$3\times 3$, $5\times 5$ and $7\times 7$ from the left to the right and from
the top to the bottom.
All the spin flips are spurious and should be disgarded; from the pictures
it is clear that
by considering blocks of different increasing sizes, one is  able to eliminate
the effects of noise. 
We see that in this case a $7 \times 7$
window is enough to label correctly  almost all the sites. 

\section{Results}
\label{sec:risultati}
Let us firstly consider the quenching at zero temperature for our
coarsening system. The fraction of spins that never flipped from
the initial time is found to decrease over time  according to the
same persistence exponent $\theta =0.22$ 
as the one estimated for the
two-dimensional Ising model [\onlinecite{DBG}]
(see Fig. \ref{fig:t=0}, the lower curve).
On the other hand one may consider the
fraction $Q(t)$ of spins that never changed phase since the
beginning of the process, and the time evolution of this quantity
is related to the probability that a given point is crossed for the
first time by a domain wall.
In order to label a site as belonging to a domain of one of
the four phases, we compared the system's configuration in a
small window centered at that site with patterns corresponding to
the four ground states, as described in Section \ref{sec:modello}.
The estimate for the persistence exponent was found to be
$\theta=0.42$.  In Fig. \ref{fig:t=0} the center and top plots represent
the logarithm of the number of persistent sites versus the logarithm of
the number of iterations in a $600\times 600$ and $1200\times 1200$
system, respectively.
We remark that this result differs from what it is observed 
in the case
of the two-dimensional four state Potts model, 
which shares with the
present model the feature of having a four-folded ground state:
as described in [\onlinecite{DOS}], the log-log plot of the number of persistent spins versus the iteration number, in the Potts case, shows some curvature and an exponent  
$\theta=0.36$ has been estimated in the case of the triangular lattice.
Moreover we note that all results on persistence exponent do not
depend on the choice of the $J's$ in the superantiferromagnetic
phase, differently from what is found with respect to 
correlation properties [\onlinecite{CGS1},\onlinecite{CGS2}].
\par
The independence from temperature of the persistence exponent has
also been checked: in Fig. \ref{fig:j2=-1} we depict the time evolution of
of the logarithm of the number of persistent sites
after a quench with $J_1=J_3=0$, $J_2=-1$ and in the cases
$l$ equal to $1\;(\bullet)$, $2\;(\blacksquare)$ and $3\;(\square)$. 
As it has been explained in
Section \ref{sec:modello}, the procedure we used to label the sites
is not sensitive to thermal noise provided that a wide enough
window is chosen. As it appears from Fig. \ref{fig:j2=-1}, the exponential
decay of $Q(t)$ is manifest, in the time interval we considered,
only in the case of $3\times 3$ windows, while using $5 \times 5$
or $7\times 7$ windows cancels the effects of temperature. 
The estimated value for $\theta$ is $\theta=0.43$,
consistent with the case $T=0$.
\par
When the temperature is increased, wider windows are needed. For
$J_1=J_3=0$ and $J_2=-0.6$ the exponential decay is manifest even
using a $7\times 7$ window, but a $9\times 9$ window is capable
to cancel the effect of noise.
In Fig. \ref{fig:j2=-06}
Monte Carlo data are plotted in cases $l$ equals 
$2\;(\bullet)$, $3\;(\blacksquare)$, $4\;(\square)$ and $5\;(\triangle)$. 
In the case $l=5$, corresponding to $11\times 11$ windows, 
we find $\theta=0.47$. A similar variation of $\theta$ near the critical temperature has been numerically observed in the Ising and Potts models  [\onlinecite{D}]; a possible mechanism for such dependence would be that the anisotropy of the surface tension in these lattice models depends on the temperature (see [\onlinecite{D}]). Better numerical simulations are needed to confirm that this variation of $\theta$ with temperature is real and due to the above described mechanism.
\par
Finally we applied
Derrida's method to evaluate the persistence
exponent for the present model at finite temperature.
We evolved two systems, one starting from a disordered and the other from a ground state, subject to the same thermal noise. The $3\times 3$ block has been used to label sites and color changes occuring simultaneously in both systems were not taken into account. 
Results similar to the ones described above are obtained.
Indeed in Fig. \ref{fig:derrida} we plot the logarithm of the number
of persistent sites, measured following Derrida's scheme, in the case
$J_1=J_3=0$ and $J_2=-1$, with lattice sizes
$256\;(\bullet)$, $600\;(\blacksquare)$ and $1200\;(\square)$.
The estimate for the persistence exponent was found not to depend on the lattice sizes we used, and it was equal to $\theta=0.43$ .

\section{Conclusions}
\label{sec:concl}
In this paper we have discussed persistence properties
of a two-dimensional spin system
with nearest neighbor, next-to-the-nearest
neighbor and plaquette interactions, 
quenched from a disordered phase into the superantiferromagnetic phase.
\par
Due the particular nature
of the ground states, the order parameter is defined in terms of 
blocks of spins. 
A procedure has been introduced to label each site of the lattice
as belonging to one of the four possible lamellar ground states.
This method allows the definition of the persistence exponent
both at zero and finite temperature, provided a large enough block
$S_l$ is used.
\par
Our estimate of the persistence exponent,
$\theta=0.42$, differs from those of the two-dimensional 
Ising and four state Potts models. 
Our results are compatible with the hypothesis that $\theta$
does not depend on temperature in the low-temperature phase of the quenching.
Some variation of $\theta$ is  observed when the temperature is close to the critical temperature, but further investigations would be needed to better characterize the persistence properties of the separating system near the critical point.

\begin{acknowledgements}
We express our thanks to Prof. B. Derrida for a valuable discussion
on the topic of this work and for useful suggestions. 
\end{acknowledgements}

\begin{figure}
\vskip +20mm
\inseps{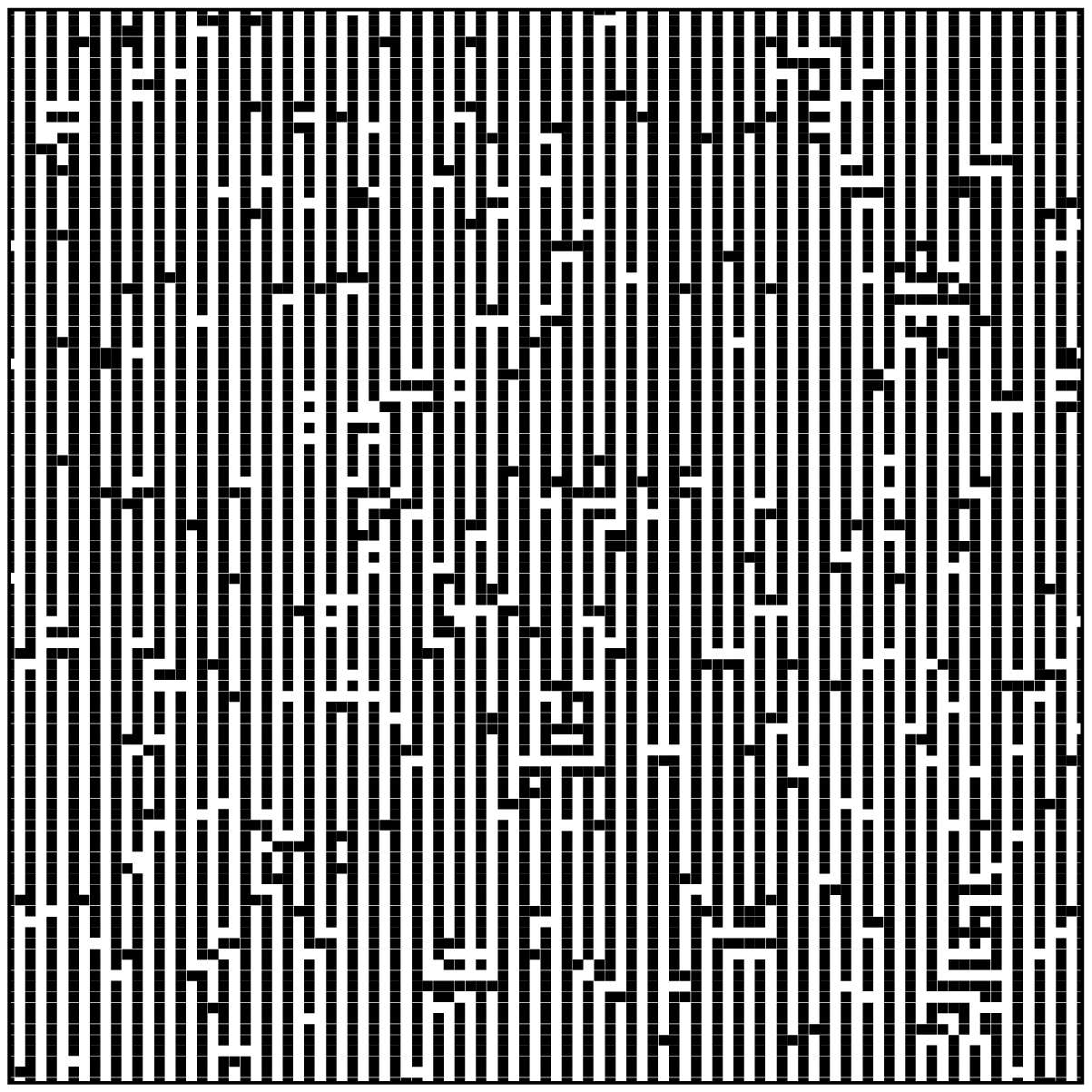}{1}
\vskip +10mm
\caption{}
\label{fig:config}
\end{figure}
\begin{figure}
\vskip -10mm
\inseps{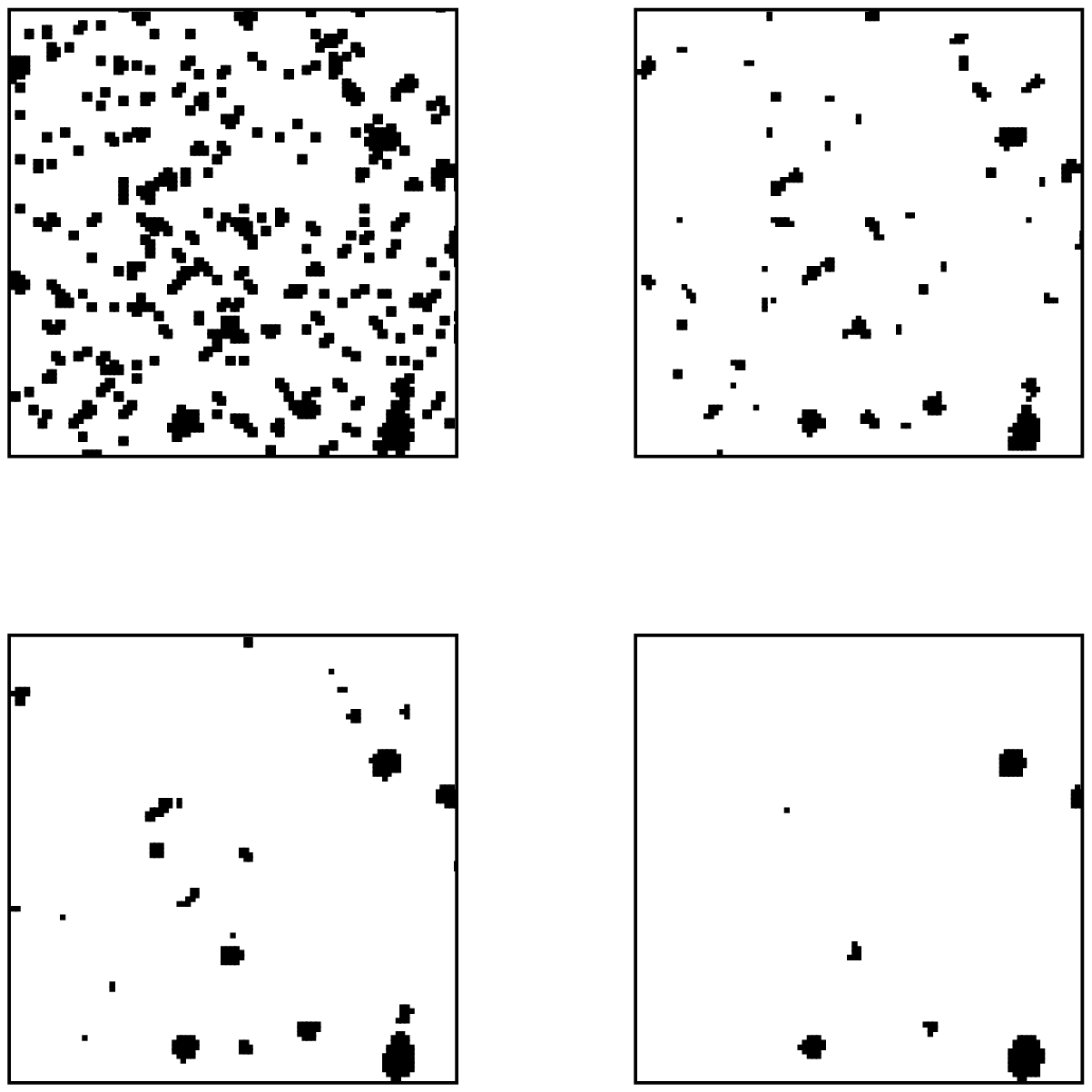}{1}
\vskip +20mm
\caption{}
\label{fig:filtri}
\end{figure}
\begin{figure}
\vskip -10mm
\inseps{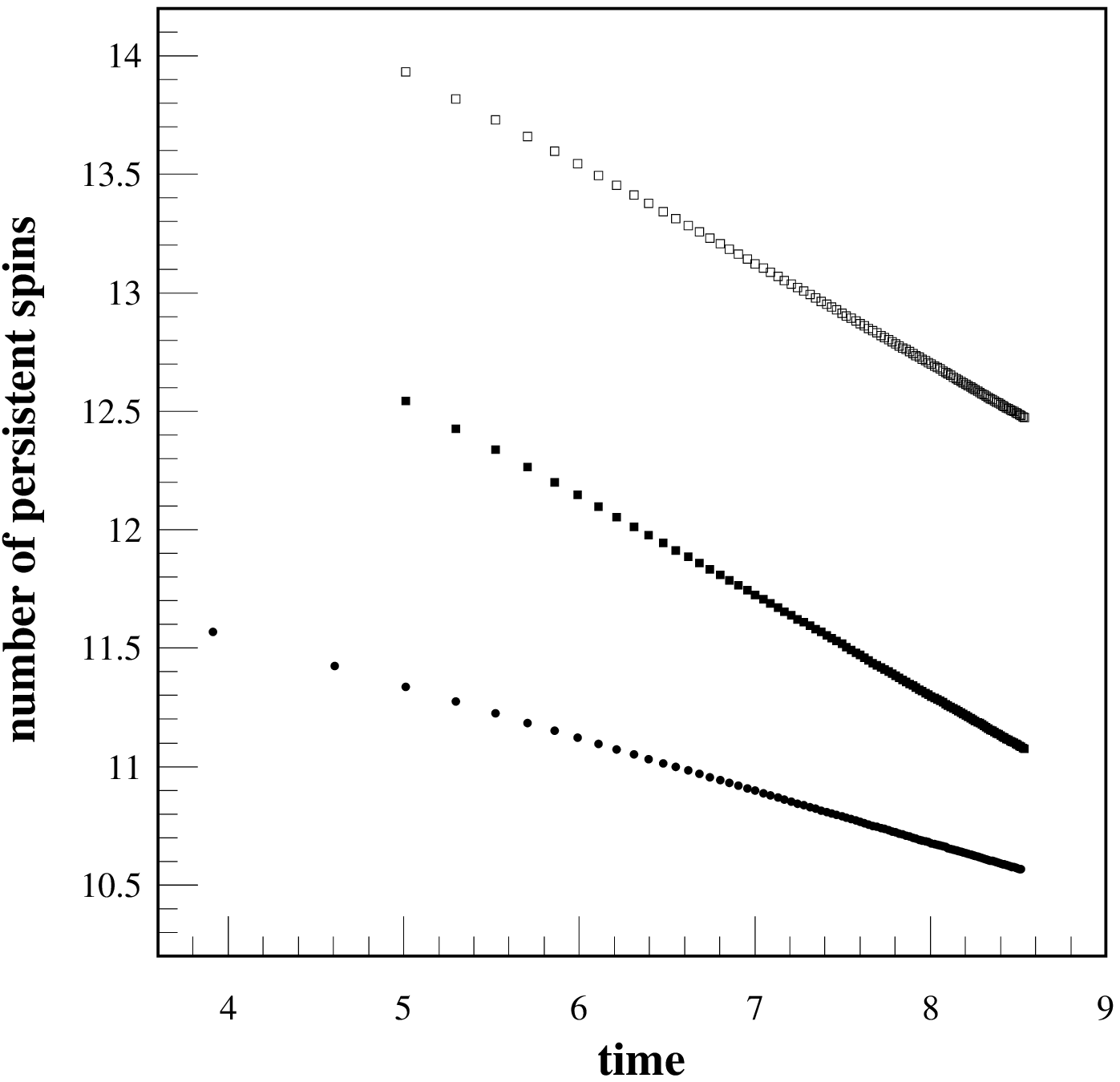}{1}
\vskip +20mm
\caption{}
\label{fig:t=0}
\end{figure}
\begin{figure}
\vskip -10mm
\inseps{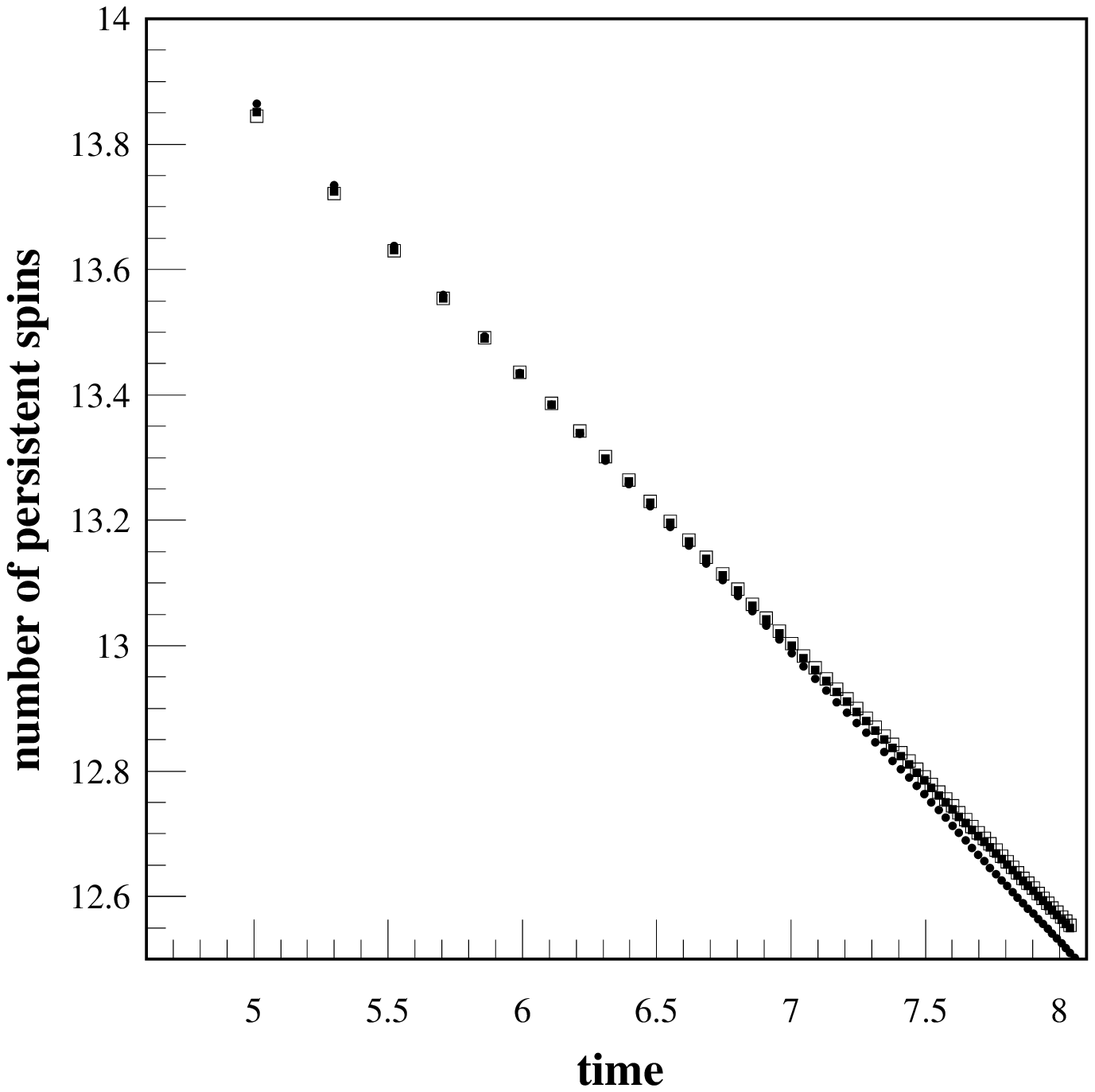}{1}
\vskip +20mm
\caption{}
\label{fig:j2=-1}
\end{figure}
\begin{figure}
\vskip -10mm
\inseps{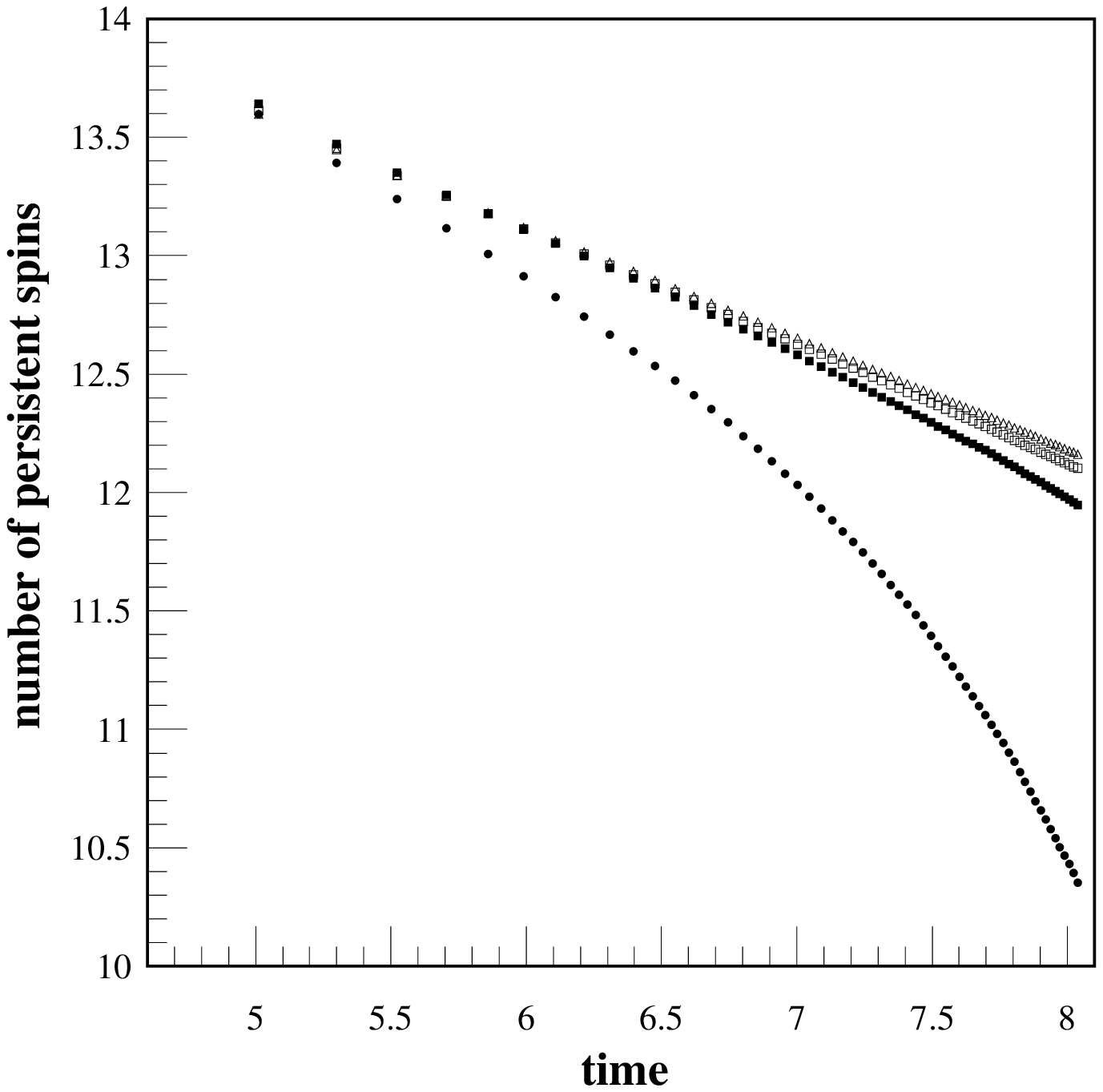}{1}
\vskip +20mm
\caption{}
\label{fig:j2=-06}
\end{figure}
\begin{figure}
\vskip -10mm
\inseps{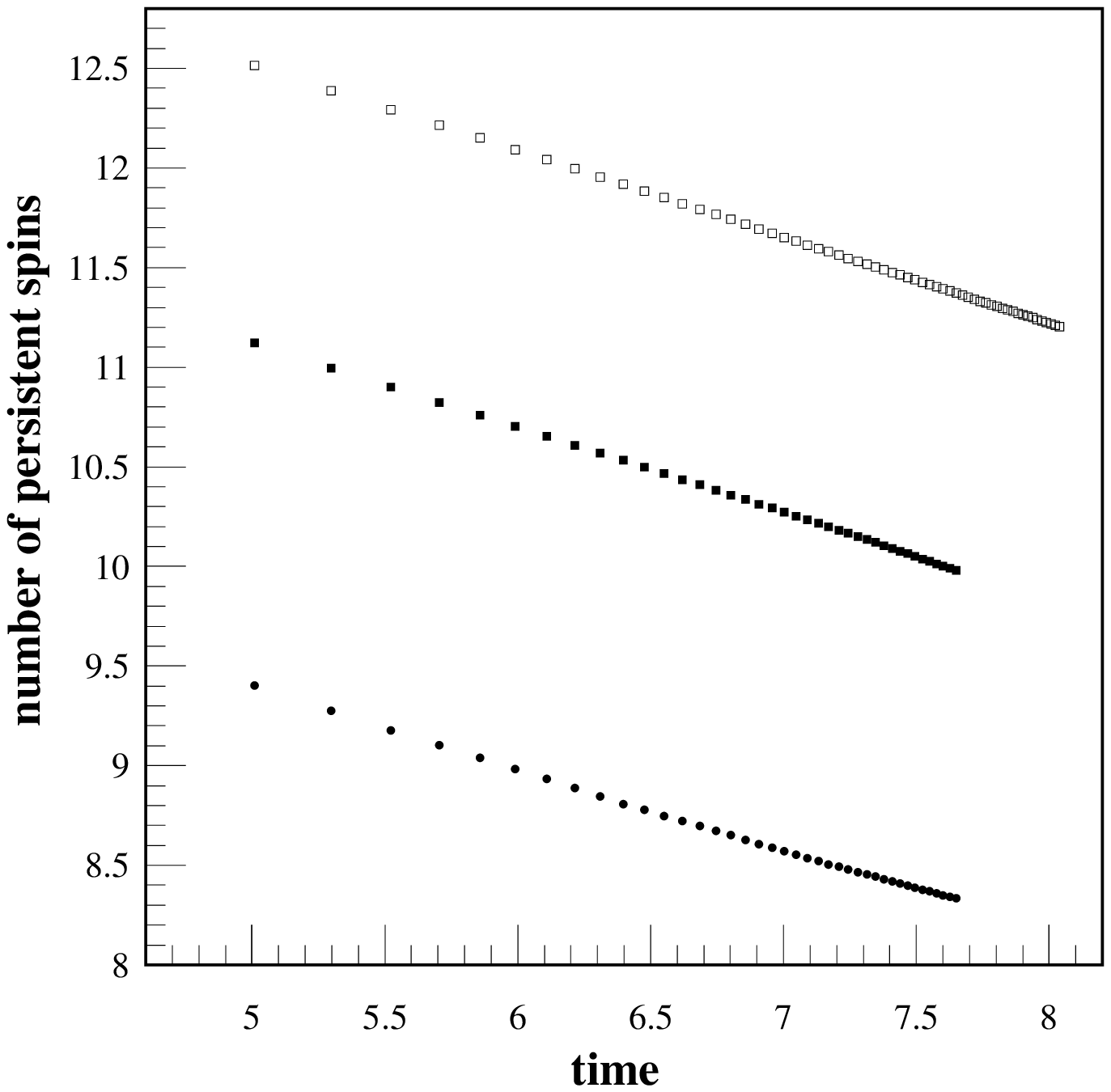}{1}
\vskip +20mm
\caption{}
\label{fig:derrida}
\end{figure}

\newpage
\centerline{\bf Figure captions}
\vskip 1 cm
\noindent{\bf Fig.1:}
Configuration of model (\ref{eq:ham}) after $1000$ full updates of
the lattice obtained starting from a completely ordered initial 
configuration, at finite temperature with $J_1=J_3=0$ and
$J_2=-0.5$. 
Black and white squares represent, respectively, plus and minus
spins.
\vskip 0.5 cm 
\noindent{\bf Fig.2:}
Starting from the configuration depicted in Fig. \ref{fig:config} 
we have applied our block procedure to label all the sites of the
lattice. Since all defects in Fig. \ref{fig:config} are thermal due,
all the lattice sites should be recognized as belonging to the same phase.
In the four pictures black squares represent sites not recognized as
belonging to the ordered phase for blocks $S_l$ $2\times 2$ 
[\onlinecite{osservazione}], 
$3\times 3$, $5\times 5$ and $7\times 7$ from the left to the right and from
the top to the bottom.
\vskip 0.5 cm 
\noindent{\bf Fig.3:}
The logarithm of the number of spins that never flipped from
the initial time versus the logarithm of the number of
iterations is depicted in the bottom plot. Monte Carlo data have been
obtained by averaging over $51$ different runs of the $600\times 600$
system.
The center (top)    
plot represents the logarithm of the number of sites that
never changed phase versus the logarithm of time; data have been
obtained by averaging over $37$ ($12$) runs of the $600\times 600$
($1200\times 1200$) system.
All data refer to the zero temperature case with parameters
$J_1=J_3=0$ and $J_2=-1$. 
\vskip 0.5 cm 
\noindent{\bf Fig.4:}
The logarithm of the number of sites that
never changed phase is plotted versus the logarithm of time; 
data have been
obtained on a $1200\times 1200$ lattice by averaging over $20$ runs, in the
finite temperature case with parameters
$J_1=J_3=0$ and $J_2=-1$. 
Different plots refer to different sizes of blocks $S_l$:
$l$ equals $1\;(\bullet)$, $2\;(\blacksquare)$ and
$3\;(\square)$.
\vskip 0.5 cm 
\noindent{\bf Fig.5:}
The logarithm of the number of sites that
never changed phase is plotted versus the logarithm of time; 
data have been
obtained on a $1200\times 1200$ lattice by averaging over $11$ runs, in the
finite temperature case with parameters
$J_1=J_3=0$ and $J_2=-0.6$. 
Different plots refer to different sizes of blocks $S_l$:
$l$ equals $2\;(\bullet)$, $3\;(\blacksquare)$,
$4\;(\square)$ and $5\;(\triangle)$.
\vskip 0.5 cm 
\noindent{\bf Fig.6:}
The logarithm of the number of sites that
never changed phase is plotted versus the logarithm of time; 
data have been
obtained by using Derrida's method [\onlinecite{D}]
in the
finite temperature case with parameters
$J_1=J_3=0$, $J_2=-1$ and $l=1$ ($3\times 3$ windows). 
Different plots refer to different lattice sizes,
$256\;(\bullet)$, $600\;(\blacksquare)$
and $1200\;(\square)$, and to averages on more than
$50$ different runs.

\newpage
$\phantom .$

\newpage
$\phantom .$

\newpage
$\phantom .$

\newpage
$\phantom .$

\newpage
$\phantom .$

\newpage
$\phantom .$

\begin{references}
\bibitem{osservazione}
In the case $2\times 2$ the parameter $K$ has been chosen equal 
to zero, that is a site is recognized as belonging to the
ordered phase if there is a perfect match between the present configuration
and the ordered one in the $2\times 2$ block with the bottom left
corner on the given site.
\end{references}
\end{document}